\documentclass[12pt]{article}

\usepackage{graphicx}

\begin{document}

%\begin{center}
%\begin{bf}
\title{A cluster density matrix for the effective field theory with correlations}
\maketitle
%\end{bf}
%\end{center}

\begin{center}
\begin{bf}
Tadeusz Balcerzak\\
\end{bf}
\vspace{5mm}
Department of Solid State Physics, University of \L\'{o}d\'{z},\\
Pomorska 149/153, 90-236 \L\'{o}d\'{z}, Poland
\end{center}

\vspace{20mm}

\begin{bf}
Abstract\\
\end{bf}
A cluster density matrix is introduced in the form suitable for the self-consistent
calculation of relevant thermodynamic averages for the Ising model with spin $S=1/2$.
On this basis, derivation of the Gibbs free-energy for the effective field theory
of Honmura and Kaneyoshi is presented.
\vspace{2cm}


\vspace{1cm}


e-mail: t\_balcerzak@uni.lodz.pl

\newpage

\section{Introduction}
The invention of the differential operator method by Honmura and Kaneyoshi
(H-K) \cite{1}
in 1979, as an approach to the Effective Field Theory (EFT), made
possible the theoretical investigations of various magnetic models on a
large scale \cite{2,3,4,5,6,7}. First of all, the method found  application in
combination with the exact Callen-Suzuki statistical identities \cite{8,9},
giving possibility for taking into consideration the correlation function.
In its simplest version, with only spin-autocorrelations taken into
account, the method is equivalent to the Zernike \cite{10}, or to 1st Matsudaira
\cite{11} approximation. The perspectives and possible development of the EFT
method has been outlined in Ref.\cite{12}. In Ref.\cite{13} the equivalence of the
differential operator method to another, so-called the integral operator
approach, has been discussed.\\

Among wide applications of the H-K differential operator method, several
important problems should be mentioned. It is, among other things, the theory
of disordered diluted alloys \cite{3, 14, 15}, random field problems \cite{15, 16},
amorphization \cite{14, 17, 18} and surface magnetism \cite{4, 19}. In the equivalent,
integral operator version, the method has been applied to studies of thin
magnetic films \cite{20, 21}. At the same time, great efforts were put into
application of H-K method to higher spins problem \cite{22, 23}. In this area, both the
pure spin models (containing the spins of the same kind), for instance the
isotropic Blume-Emery-Griffiths model \cite{24}, as well as the mixed-spin
systems \cite{6, 25} have been considered. A lot of attention has also been paid
to the transversal models \cite{7, 26, 27}.\\

Despite wide applications and almost 30 year long history of the method,
the Gibbs free-energy in H-K approximation has not been found so far. Some attempts
in this direction \cite{28, 29} cannot be considered
as satisfactory because they fail to lead to the self-consistent
thermodynamics. The only thermodynamic potential which has been correctly
defined up to now is the internal energy \cite{30}.\\

On the other hand, Gibbs free-energy is of particular importance for the thermodynamic description. 
Not least because in case of the 1st order phase transitions, or when the several
mathematical solutions can occur, one has to have a clear criterion to
choose up the physical solution corresponding to the
minimum of free energy. Such problems often occur in connection with
the structural disorder in the system, especially in case of higher spins
$(S>1/2)$ and the presence of anisotropy.\\

The aim of the present paper is a construction of the Gibbs energy 
in H-K approximation, starting from a new form proposed for the cluster density matrix. 
In order to describe the method in detail, the simplest case, i.e., the crystalline Ising model with
spin $S=1/2$, is considered. Further progress of the proposed approach (for
instance, implementation for higher spins) should be possible on this basis.
In the derivation of the Gibbs free-energy we demand that
some known EFT equations for the magnetization and NN correlation
function remain fulfilled.\\

Among the results, the formulas for the entropy and Gibbs free-energy in H-K approximation
are presented. Some testing numerical calculations illustrate the Gibbs free-energy behaviour upon the
temperature and, by comparison with other mean-field results, serve as a control for the presented formalism.\\

\section{The cluster density matrix}
Below, we  consider the crystalline Ising model in an arbitrary
dimension with spin $S=1/2$. We take into account the standard
nearest-neighbour (NN) exchange interaction and the external field terms.
Hamiltonian can be written in the form:
%1
\begin{equation}\label{eq1}
\mathcal{H} =-J\sum\limits _{\langle i,j \rangle} S_{i}S_{j} -h \sum\limits
_{j} S_{j} \end{equation}

\noindent
where $S_{i}$, $S_{j}=\pm 1/2$ are the $z$-components of the spins in
$i$-th and $j$-th site, respectively, $J$ is the exchange interaction for
NN, and $h=-g \mu_{\rm B}H$ stands for an external magnetic field $H$
oriented in $z$-direction.\\

In the system we  distinguish a $(z+1)$-atomic cluster, consisting of
a central spin $S_{i}$ and $z$-spins $S_{j}$ being NN of the central spin.
The cluster is embedded in an external magnetization field.
We restrict our considerations for those lattices where the spins
$S_{j}$ in the neighbourhood of $S_{i}$ are not the mutual NN. 
This excludes, for instance, the triangular and FCC lattices. However, it should be mentioned that in the case of triangular lattice, the stair-triangle transformation \cite{31,32} can be applied for mapping onto honeycomb lattice (and vice-versa).\\

The density
matrix for such a cluster can be proposed in the following factorial
form:
%2
\begin{equation}\label{eq2}
\rho _{i\{  NN\}} =\left\{ \frac{1}{2} +S _{i} \tanh \left[ \frac{\beta}{2} \left(
J \sum\limits _{j \in i}^{z} S _{j} +h\right) \right] \right\} \cdot \prod\limits_{j
\in i}^{z} \left( \frac{1}{2} +2m S_{j} \right) \end{equation}

\noindent
where $m\stackrel{\textit{\scriptsize def}}{=}\langle S_{i}\rangle=\langle S_{j}\rangle$
is a self-consistent
parameter being the thermodynamic mean value of a given spin from the
cluster and corresponding to the local magnetization. 
The form of $\rho_{i\{ NN\}}$ can be interpreted as an outer product of $2 \times 2$ sigle-site matrices, giving $2^{z+1} \times 2^{z+1}$ diagonal matrix for the cluster.
It is easy to check
that the cluster density matrix (\ref{eq2}) satisfies the normalization condition:
%3
\begin{equation}\label{eq3}
\textrm{Tr} _{i\{ NN\}}\rho  _{i\{ NN\}}=1
\end{equation}

\noindent
where the trace $\textrm{Tr}_{i\{NN\}}$ is performed over the central $i$-th
spin and all the  NN spins.\\

The cluster density matrix (\ref{eq2}) can be used for calculation of the local magnetization $m$:
%4
\begin{equation}\label{eq4}
m=\textrm{Tr} _{i\{ NN\}}\left( S_{i}\rho _{i\{ NN\}} \right)
\end{equation}

\noindent
and the NN correlation function $c$, defined as:
$c\stackrel{\textit {\scriptsize def}}{=}\langle S_{i}S_{j} \rangle$, namely:
%5
\begin{equation}\label{eq5}
c=\textrm{Tr} _{i\{ NN\}} \left( S_{i}S_{j}\rho _{i\{ NN\}} \right)
\end{equation}

\noindent
For convenience of calculations of $m$ and $c$ let us note that the matrix $\rho_{i\{NN\}}$
can be partially reduced.
For instance, performing the trace over all NN-spins  we obtain the single-site
density matrix $\rho_{i}$
%6
\begin{equation}\label{eq6}
\rho_{i} =\textrm{Tr}_{\{ NN \}} \rho  _{i\{ NN\}} =\frac{1}{2} +2 \left( a_{0}+ a_{1}m
+ \ldots + a _{z}m^{z} \right)S_{i}
\end{equation}

\noindent
where $a_{0}, ... ,a_{z}$ are the temperature and field dependent coefficients, whose
explicit form for a given $z$-number are presented in the Appendix.\\

On the other hand, the single-site density matrix can be generally presented
in the form (see e.g.Ref.\cite{33}):
%7
\begin{equation}\label{eq7}
\rho_{i}=\frac{1}{2} + 2mS_{i}
\end{equation}

\noindent
Hence, by comparison of eqs.\,(\ref{eq6}) and (\ref{eq7}) we obtain:
%8
\begin{equation}\label{eq8}
m=a_{0} +a_{1}m+\ldots +a_{z}m^{z}
\end{equation}

\noindent
which is the EFT equation for the magnetization in H-K approximation \cite{1}. This equation has
previously been derived in a different way, from the exact
Callen-Suzuki identity (see, for instance, in Refs.\cite{1,13}) through the operator techniques.\\
Another partial reduction of $\rho_{i\{NN\}}$ given by eq. (\ref{eq2}) can be obtained by performing the trace
over $(z-1)$-spins from NN, with
except of one $j$-th spin. This procedure leads to the pair density matrix
$\rho _{ij}$, namely:
%9
{\setlength\arraycolsep{1pt}
\begin{eqnarray}\label{eq9}
\rho_{ij} =\textrm{Tr} _{\{ NN \neq j \}} \rho _{i\{ NN\}} &=&\frac{1}{4}
+\left( a_{0}+a_{1}m+\ldots + a_{z}m^{z} \right)S_{i}+ mS_{j}\nonumber \\
&& + 4 \left( b_{0}+b_{1}m +\ldots +b_{z}m^{z} \right) S_{i}S_{j}
\end{eqnarray}}
\noindent
In (\ref{eq9}) $b_{0}, ... ,b_{z}$ are the coefficients depending on the $z$-number, and are
given in the Appendix. On the
other hand, the pair density matrix should have the following symmetrical form \cite{33}:
%10
\begin{equation}\label{eq10}
\rho_{ij} = \frac{1}{4} +mS_{i}+mS_{j}+4cS_{i}S_{j}
\end{equation}
\noindent
Hence, by comparison of eq.\,(\ref{eq9}) and (\ref{eq10}) we obtain:
%11
\begin{equation}\label{eq11}
c = b_{0}+b_{1}m+ \ldots + b_{z}m^{z}
\end{equation}
\noindent
together with eq.\,(\ref{eq8}). Eq.\,(\ref{eq11}) reproduces the EFT result in H-K approximation
for the NN correlation function, obtained from the exact
Callen-Suzuki identity (see, for instance, Refs.\cite{13, 30}). Thus, we see that some basic
EFT equations for $m$ and $c$ which are present in literature can now be  alternatively re-derived from
reducibility of the cluster density matrix $\rho_{i\{NN\}}$ (\ref{eq2}).\\

\section{The cluster entropy}
Calculation of the entropy presents usually the most difficult task in
statistical thermodynamics and has not been done yet in the H-K approximation.
In order to calculate the cluster entropy we
will make use of the cluster density matrix $\rho_{i\{NN\}}$. However, in this case the
matrix should be transformed first to a more convenient, exponential form.
For this purpose, we will use the exact identity for spin $S=1/2$:
%12
\begin{equation}\label{eq12}
e^{\beta \Lambda S_{j}} =\cosh \left( \frac{\beta \Lambda}{2} \right)+
2S_{i} \sinh \left( \frac{\beta \Lambda}{2} \right) \end{equation}

\noindent
With the help of (\ref{eq12}) we obtain the next useful relationship:
%13
\begin{equation}\label{eq13}
\frac{1}{2} + 2mS_{j} =\frac{e^{\beta \Lambda S_{j}}}{2\cosh\left(
\frac{\beta \Lambda}{2} \right)} \end{equation}

\noindent
where
%14
\begin{equation}\label{eq14}
m=\frac{1}{2} \tanh\left( \frac{\beta \Lambda}{2} \right)
\end{equation}

\noindent
Eq.\,(\ref{eq14}) defines the $\Lambda$-parameter, which can be unambigously
determined when the magnetization $m$ is calculated from eq.\,(\ref{eq8}). With the
help of (\ref{eq13}) the cluster density matrix (\ref{eq2}) can be presented in the 
exponential form:
%15
\begin{equation}\label{eq15}
\rho _{i\{ NN\}} =\frac{\exp \left[ \beta\left( J\sum\limits_{j \in i}^{z} S_{j}+h
\right)S_{i}\right]}{2 \cosh \left[ \frac{\beta}{2}\left( J\sum\limits _{j \in i}^{z}
S_{j}+h \right) \right]} \times \frac{\exp \left(\beta \Lambda \sum\limits
_{j \in i}^{z} S_{j}\right)}{2^{z} \cosh ^{z} \left( \frac{\beta \Lambda}{2}
\right)}\end{equation}

\noindent
As a next step, the cluster entropy $\sigma_{i\{NN\}}$
can be calculated from the formula:
%16
\begin{equation}\label{eq16}
\sigma _{i\{ NN\}} =-k _{\rm B} \langle \ln \rho _{i\{ NN\}} \rangle =-k_{\rm B}
\textrm{Tr} _{i\{ NN\}} \left( \rho _{i\{ NN\}} \ln \rho _{i\{ NN\}} \right)
\end{equation}

\noindent
where $\rho_{i\{NN\}}$ is given by (\ref{eq15}). As a result we obtain:
%17
{\setlength\arraycolsep{1pt}
\begin{eqnarray}\label{eq17}
\sigma _{i\{ NN\}} &=&-\frac{1}{T} \left( zJc +hm+z\Lambda m \right) +zk
_{\rm B} \ln \left[ 2\cosh \left( \frac{\beta \Lambda}{2} \right)
\right]\nonumber\\
&& + k_{\rm B} \left\langle \ln \left\{ 2 \cosh \left[ \frac{\beta}{2} \left( J
\sum\limits _{j \in i}^{z} S_{j}+h \right) \right] \right\} \right\rangle
\end{eqnarray}}

\noindent
The average value of the logarithm function remaining in the last term of eq.\,(\ref{eq17}) can be conveniently
calculated with the help of
$\rho_{i\{NN\}}$ in the form of (\ref{eq2}). The final result is a polynomial of $m$:
%18
\begin{equation}\label{eq18}
\left\langle \ln \left\{ 2\cosh \left[ \frac{\beta}{2} \left( J\sum\limits _{j \in
i}^{z} S_{j}+h \right) \right] \right\} \right\rangle  = c_{0} + c_{1}m + \ldots +
c_{z}m^{z} \end{equation}

\noindent
where $c_{0}, ... ,c_{z}$  coefficients are given in the Appendix. Finally,
the cluster entropy (\ref{eq17}) is given by the equation:
%19
{\setlength\arraycolsep{1pt}
\begin{eqnarray}\label{eq19}
T\sigma _{i\{ NN\}} & = & -zJc-hm-z\Lambda m +zk_{\rm B}T\ln \left[ 2 \cosh
\left( \frac{\beta \Lambda}{2} \right) \right]\nonumber \\ && + k_{\rm B}T
\left( c_{0} + c_{1}m + \ldots + c_{z}m^{z} \right) \end{eqnarray}}

\noindent
It is seen from eq.\,(\ref{eq19}) that the cluster entropy can be obtained for a
given structure ($z$-number) when $m$ and $c$ are calculated
first as the solutions of EFT equations (\ref{eq8}) and (\ref{eq11}), respectively.\\

\section{The Gibbs free-energy}
As stated before, a correct Gibbs potential has not been obtained yet in the H-K method.
Here we postulate that the Gibbs free-energy of a small cluster interacting with its neighbourhood 
and with an external field can be written in the form:
%20
\begin{equation}\label{eq20}
G _{i\{ NN\}} = G_{0} -zJc -zm\lambda -\left( z+1 \right)hm -T\sigma _{i\{ j\}}
\end{equation}

\noindent
where $G_{0}$ is some constant part of the internal energy ( which doesn't depend
 on the temperature), $-zJc$ is the internal energy of the cluster
connected with the internal NN corelations only, and $\lambda$-parameter
corresponds to the effective-field interaction of the cluster boundary with the rest of the
system. In eq.\,(\ref{eq20}) we have also the Zeeman term $-(z+1)hm$, as well as the
entropic term $-T\sigma_{i\{NN\}}$, given by eq.\,(\ref{eq19}). Then, substituting eq.\,(\ref{eq19})
into (\ref{eq20}) the cluster Gibbs free-energy can be written as:
%21
{\setlength\arraycolsep{1pt}
\begin{eqnarray}\label{eq21}
G _{i\{ NN\}} & = & G_{0} -zm (\lambda + \Lambda +h) -zk_{\rm B}T \ln \left[ 2
\cosh \left( \frac{\beta \Lambda}{2} \right) \right]\nonumber \\
&& - k_{\rm B}T \left( c_{0} +c_{1}m +\ldots + c_{z}m^{z} \right)
\end{eqnarray}}

\noindent
where $\Lambda$ is given by eq.\,(\ref{eq14}). The new parameter $\lambda$ can now be
determined from the condition that the Gibbs energy in equilibrium should
be minimized with respect to it, or equivalently, $G_{i\{NN\}}$ should be in
a minimum with respect to the magnetization, since ${\partial
G_{i\{NN\}}}/{\partial \lambda}= \left({\partial G_{i\{NN\}}}/{\partial m}\right)\left(
{\partial m}/{\partial \lambda}\right)$. Thus we demand that:
%22
\begin{equation}\label{eq22}
\frac{\partial G _{i\{ NN\}}}{\partial m} =0 .
\end{equation}

\noindent
Eq.\,(\ref{eq22}) presents the necessary extremum condition for the free-energy, exploited in various effective field approaches \cite{28,33}. 
In our case, from this condition we obtain the equation for $\lambda$,
namely:
%23
\begin{equation}\label{eq23}
m \frac{\partial \lambda}{\partial m}+ \lambda =\Lambda -h-\frac{1}{z}
k_{\rm B}T \left( c_{1} +2c_{2}m + \ldots +zc_{z}m^{z-1} \right)
\end{equation}

\noindent
The above differential equation can be numerically solved only when the
magnetization $m$ is calculated first from eq.\,(\ref{eq8}). The boundary condition
for $\lambda$ can easily be established for the
paramagnetic state. Namely, if we put $h=0$, $m=0$ and
$\Lambda=0$, then from (\ref{eq23}) we obtain $\lambda=0$. Thus, we can assume as a
boundary condition that $\lambda=0$ above the Curie
temperature.\\

After calculations of $\lambda$ vs. temperature
the constant $G_{0}$ in eq.\,(\ref{eq20}) can be found from the ground state
energy. Namely, we should demand that the internal energy per site in the ground state (at $T=0$)
when calculated from (\ref{eq20}) is exact, as it results from states of the hamiltonian (\ref{eq1}). Thus
we have the condition:
%24
\begin{equation}\label{eq24}
\left( \frac{G _{i\{ NN\}}}{z+1} \right) _{T=0 \atop h=0} =
\frac{G_{0}-\left( \frac{1}{4}J+\frac{1}{2} \lambda _{0}  \right)z}{z+1} =
\left( \frac{\langle \mathcal{H} \rangle}{N}  \right)_{T=0 \atop h=0}
=-\frac{Jz}{8} \end{equation}

\noindent
where $\lambda_{0}=\lambda (T=0)$. Solving eq.\,(\ref{eq24}) we determine the constant:
%25
\begin{equation}\label{eq25}
G_{0} =\frac{z}{2} \left[ \lambda _{0} -\frac{J}{4} \left( z-1 \right)
\right] \end{equation}

\noindent It is worth noticing that the Gibbs energy
(\ref{eq21}) is fully described in terms of $h$ and $T$ as the primary variables,
whereas the intermediate variable $m(h,T)$ should be treated as a solution of eq.\,(\ref{eq8}).\\

It is necessary to check whether the Gibbs energy derivatives properly
reproduce basic thermodynamic quantities.
Differentiating eq.\,(\ref{eq21}) over $h$, with the help of (\ref{eq23}), we obtain after calculation:
%26
\begin{equation}\label{eq26}
\left( \frac{\partial G _{i\{NN\}}}{\partial h} \right)_{T} =-(z+1)m
\end{equation}

\noindent
which gives the cluster magnetization with $m$ in the form (\ref{eq8}). During the differentiation we have used
the following relationship valid for the
coefficients given in the Appendix:
%27
\begin{equation}\label{eq27}
\left( \frac{\partial c_{n}}{\partial h} \right)_{T} =\beta a _{n}
\end{equation}

On the other hand, by differentiation of eq.\,(\ref{eq21}) over $T$ (with the help of (\ref{eq23}))
we are able to reproduce the cluster entropy:
%28
\begin{equation}\label{eq28}
\left( \frac{\partial G_{i\{NN\}}}{\partial T} \right)_{h} =-\sigma_{i\{NN\}}
\end{equation}

\noindent
where $\sigma_{i\{NN\}}$ is given by eq.\,(\ref{eq19}). In this differentiation we made
use of another useful relationship
for the coefficients:
%29
\begin{equation}\label{eq29}
\left( \frac{\partial c_{n}}{\partial T} \right)_{h}
=-\frac{1}{k_{\rm B}T^{2}} \left( a_{n}h+zJb_{n} \right) \end{equation}

In a consequence of the first derivatives check, the second derivatives of
the Gibbs potential (\ref{eq21}) will lead to the expressions for the
isothermal susceptibility $\chi _{T}$ and the magnetic specific heat at
constant $h$, $C_{h}$, respectively:
%30
\begin{equation}\label{eq30}
\chi _{T}=-\left( \frac{\partial^{2}G_{i\{NN\}}}{\partial h^{2}} \right)_{T}
=(z+1) \left( \frac{\partial m}{\partial h} \right)_{T} \end{equation}

\noindent
and
%31
\begin{equation}\label{eq31}
C_{h} =-T\left( \frac{\partial^{2}G_{i\{NN\}}}{\partial T^{2}} \right)_{h} =
T \left( \frac{\partial \sigma _{i\{NN\}}}{\partial T} \right) _{h}
\end{equation}

\noindent
Moreover, for the independent variables $h$ and $T$, the mixed derivatives of the Gibbs potential must satisfy the identity:
%32
\begin{equation}\label{eq32}
\frac{\partial^{2} G_{i\{ NN\}}}{\partial T \partial h} 
= \frac{\partial^{2} G_{i\{ NN\}}}{\partial h \partial T}         
\end{equation}

\noindent
Taking into account that eqs. (\ref{eq26}) and (\ref{eq28}) are fulfilled, we can write eq. (\ref{eq32}) in the form of Maxwell relation:
%33
\begin{equation}\label{eq33}
\left( z+1 \right)\left( \frac{\partial m}{\partial T} \right)_{h}
= \left( \frac{\partial \sigma_{i\{ NN\}}}{\partial h} \right)_{T}    
\end{equation}

\noindent
In practice, the validity of relation (\ref{eq33}) can be checked for a given $z$, $h$ and $T$ numerically only, because the analytical expresions for $m$ and $\sigma_{i\{ NN\}}$ (eqs.(\ref{eq8}) and (\ref{eq19}), respectively) are too complex. \\

The equations (\ref{eq26}, \ref{eq28}, and \ref{eq30}, \ref{eq31}) allow us to study the basic properties of the model from the free-energy expression, whereas
the equilibrium condition (\ref{eq22}), and the existing EFT equations in H-K approximation (\ref{eq8}, \ref{eq11}), remain fulfilled.\\

\section{Discussion of the results and conclusion}
In the paper a method of derivation of the Gibbs free-energy
within H-K approximation has been shown. In calculations one has to solve the differential
equation (\ref{eq23}) for the variational parameter $\lambda$. In order to perform
a numerical test for the theory we have calculated the $\lambda$-parameter vs. temperature
for $h=0$ and $z=4$. The testing results are gathered in
Fig.\,1. For comparison, the magnetization $m$ and NN correlation
function $c$ are also presented, being the solutions of eqs.\,(\ref{eq8}) and (\ref{eq11}),
respectively. The numerical solution of differential
eq.\,(\ref{eq23}) can be done by iteration procedure. For
this purpose, it is convenient to start from the Curie point (when $\lambda
=0$) and to decrease the temperature in small steps, for each step
calculating successive $m$ and $\lambda$ until the ground state. For $T=0$
we obtained the value $\lambda_{0}=0.6261$. Then, having $\lambda$ upon
$T$, as well as $m$, $c$, and $\Lambda$ from eq.\,(\ref{eq14}), the Gibbs free-energy
can be found from eq.\,(\ref{eq21}) for arbitrary temperature.\\

\begin{figure}[h]
\begin{center}
\includegraphics[width=10cm]{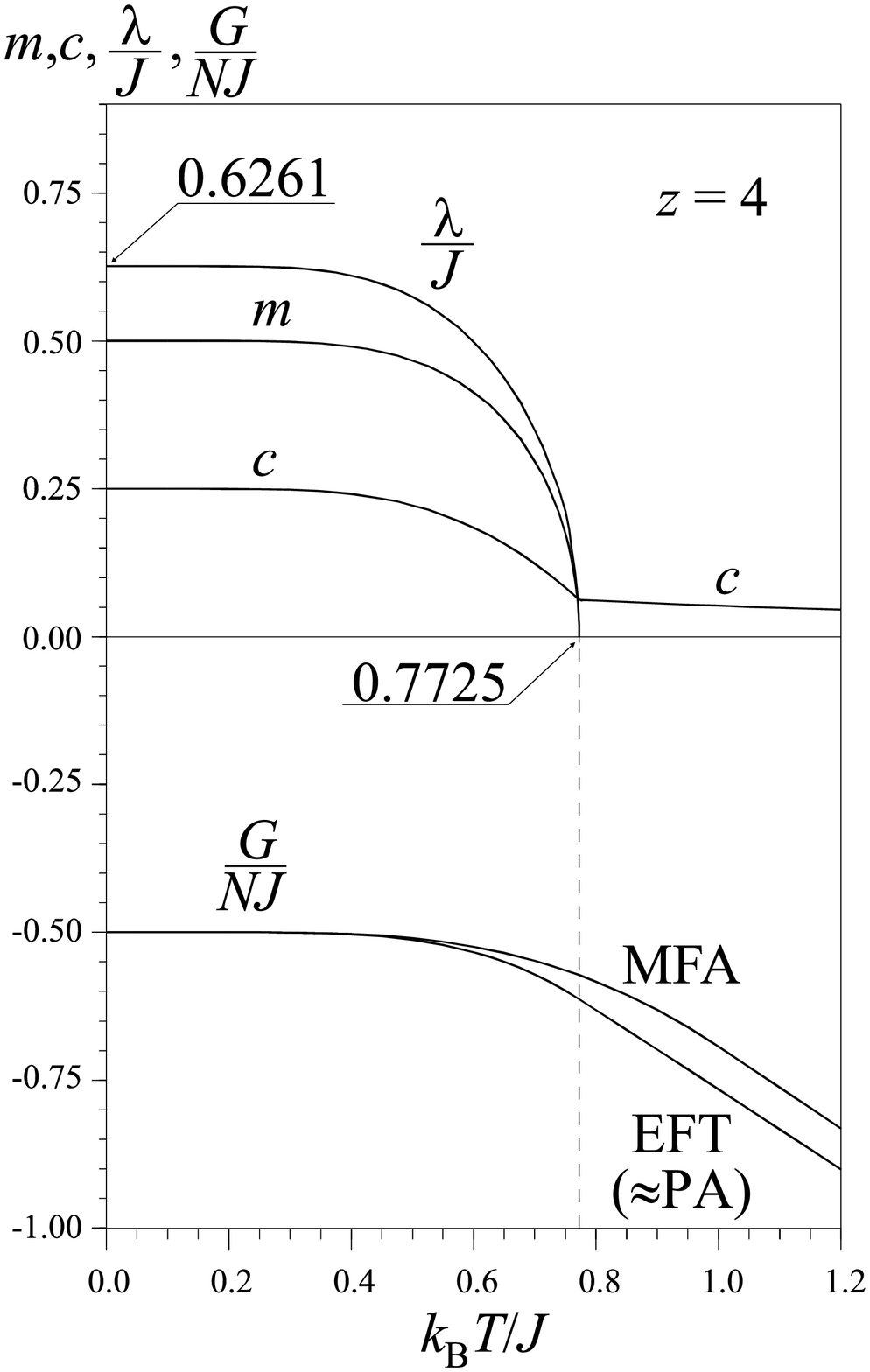}
\caption{Illustration of the variational parameter $\lambda / J$, magnetization $m$, NN correlation function $c$ and the chemical potential $G/NJ$ upon the reduced temperature $k_{\rm B}T/J$ for $z=4$ and $h=0$.}
\label{fig:1}
\end{center}
\end{figure}

The results of the Gibbs energy calculations per 1 lattice site (i.e. the
chemical potential) are also presented in Fig.\,1 for $z=4$. For comparison,
the chemical potential in the Molecular Field Approximation (MFA) and
in the Pair Approximation (PA) is also shown. The calculations for
MFA and PA are based on the theory given in Ref.\cite{33}. It turned out that the
curves for EFT in H-K approximation and PA are almost the same (within the accuracy of the line
thickness) for the range of temperatures discussed. 
This is a worth-noticing result, because the PA method is usually considered as a more accurate approach,
among other things, giving Bethe result for the Curie temperature. In fact, Curie
temperatures for $z=4$ resulting from EFT (in H-K approx.) and PA are different and they are $k_{\rm
B}T_{\rm C}/J=0.7725$ and  $k_{\rm B}T_{\rm C}/J=1/(2{\ln}2)=0.7213$,
respectively (for comparison, $k_{\rm B}T_{\rm C}/J=1$ in MFA). Explanation of this difference is due to the different 
magnetization curves (i.e. the $h$-dependence of the Gibbs energy) in both models.\\

As far as the simple cubic lattice is concerned ($z=6$), the Curie temperature resulting from EFT for $S^{z}=\pm 1/2$ is equal to $k_{\rm B}T_{\rm C}/J = 1.268$. This value corresponds to the first Matsudaira's approximation \cite{11}, and is much better then MFA result $k_{\rm B}T_{\rm C}/J = 1.500$. On the other hand, Bethe analytical result (equivalent to PA \cite{33}) is $k_{\rm B}T_{\rm C}/J = 1/(2 {\rm ln}(z/(z-2))$, which for $z=6$ gives the value $k_{\rm B}T_{\rm C}/J = 1.233$. One of the best results for the Curie temperature, $k_{\rm B}T_{\rm C}/J = 1.145$, has been obtained by Kikuchi \cite{34} by means of cluster variational method in cubic-cell approximation. The result of Kikuchi is very close to HTSE Curie temperature $k_{\rm B}T_{\rm C}/J = 1.128$ \cite{35}, which is probably the most accurate  value for s.c. lattice. The Kikuchi method is worth-mentioning because it also enables calculation of the entropy, however, by a more complex combinatorial approach.
A disadvantage of the Kikuchi method is such that it involves many variational parameters, according to which the chemical potential should be minimized. In contrast, the advantage of the present EFT theory is that only one kind of parameter (magnetization) can be chosen to minimize the Gibbs energy. Contrary to PA \cite{33}, or Kikuchi metod \cite{34}, in this formulation of EFT the correlation $c$ is not a variational parameter, but serves only as a complementary quantity being fully determined by independent variables $h$ and $T$. This simplicity of EFT provides much wider perspectives of application at the expense of accuracy. It should be stressed that EFT is superior to MFA, where the correlations are totally neglected.\\

The method presented in the paper enables complete thermodynamic description from one root expresion proposed for the cluster density matrix.
From the reducibility of this matrix, the EFT equations for magnetization and NN correlation
functions are re-derived. 
The expression for the Gibbs energy in H-K approximation is obtained more correctly than that presented in Refs.\cite{28,29}.
Let us remind that in the existing EFT approaches to the Gibbs energy the eqs. (\ref{eq26}) and (\ref{eq22}) were not simultaneously fulfilled.
In the present approach, all thermodynamic quantities can be calculated as
the successive derivatives of the Gibbs free-energy, together with the extremum condition (\ref{eq22}).\\

We hope that the method can be employed for more complicated situations, for
instance, for disordered materials (with site and/or bond disorder) as well as
for thin films and surfaces. 
As far as higher spins $(S>1/2)$ are concerned, the
problem becomes more complex because higher
spin moments are then involved in all relevant thermodynamic expressions.
However, the possible implementations of the method should be published elsewhere.\\

%\section*{Acknowledgements}

\section{Appendix:\\ A set of coefficients for $z=2$, $4$ and $6$} 

\setcounter{equation}{0}
\renewcommand{\theequation}{A\arabic{equation}}

We define the functions:
%A1
\begin{equation}
\left\{ \begin{array}{l}  F_{n} =\tanh \left[ \frac{\beta}{2}\left( nJ+h
\right) \right] - \tanh \left[ \frac{\beta}{2}\left( nJ-h \right) \right] \\
G_{n} =\tanh \left[ \frac{\beta}{2}\left( nJ+h
\right) \right] + \tanh \left[ \frac{\beta}{2}\left( nJ-h \right) \right]
\end{array} \right. \end{equation}

\noindent and
%A2
\begin{equation}
\left\{ \begin{array}{l}  X_{n} =\ln\left\{2\cosh \left[
\frac{\beta}{2}\left( nJ+h \right) \right]\right\} - \ln\left\{2\cosh \left[
\frac{\beta}{2}\left( nJ-h \right) \right]\right\} \\
Y_{n} =\ln \left\{2\cosh \left[
\frac{\beta}{2}\left( nJ+h \right) \right]\right\} + \ln \left\{2\cosh \left[
\frac{\beta}{2}\left( nJ-h \right) \right]\right\} \end{array} \right.
\end{equation}

\noindent The temperature and field dependent coefficients from eqs.\,(\ref{eq8}),
(\ref{eq11}) and (\ref{eq18}) are the following:

\noindent $z=2$ case:
%A3
{\setlength\arraycolsep{1pt}
\begin{eqnarray}
a_{0}&=& \frac{1}{8} \left( F_{1}+F_{0} \right) \nonumber \\
a_{1} & = & \frac{1}{2} G_{1}\\
a_{2} &=& \frac{1}{2} \left( F_{1} -F_{0} \right) \nonumber
\end{eqnarray}}
%A4
{\setlength\arraycolsep{1pt}
\begin{eqnarray}
b_{0}&=& \frac{1}{16} G_{1} \nonumber \\
b_{1} & = & \frac{1}{4} F_{1}\\
b_{2} &=& \frac{1}{4} G_{1} \nonumber
\end{eqnarray}}
%A5
{\setlength\arraycolsep{1pt}
\begin{eqnarray}
c_{0}&=& \frac{1}{4} \left( Y_{1}+Y_{0} \right) \nonumber \\
c_{1} & = & X_{1}\\
c_{2} &=& Y_{1} -Y_{0} \nonumber
\end{eqnarray}}

\noindent $z=4$ case:
%A6
{\setlength\arraycolsep{1pt}
\begin{eqnarray}
a_{0}&=& \frac{1}{32} \left( F_{2}+4F_{1}+3F_{0} \right) \nonumber \\
a_{1} & = & \frac{1}{4} \left(G_{2}+2 G_{1} \right) \nonumber \\
a_{2} &=& \frac{3}{4} \left( F_{2} -F_{0} \right) \\
a_{3} &=& \left( G_{2}-2G_{1} \right) \nonumber \\
a_{4} &=& \frac{1}{2} \left( F_{2} -4F_{1}+3F_{0} \right) \nonumber
\end{eqnarray}}
%A7
{\setlength\arraycolsep{1pt}
\begin{eqnarray}
b_{0}&=& \frac{1}{64} \left( G_{2}+2G_{1} \right) \nonumber \\
b_{1} & = & \frac{1}{8} \left(F_{2}+F_{1} \right) \nonumber \\
b_{2} &=& \frac{3}{8}G_{2}  \\
b_{3} &=& \frac{1}{2}\left( F_{2}-F_{1} \right) \nonumber \\
b_{4} &=& \frac{1}{4} \left( G_{2} -2G_{1} \right) \nonumber
\end{eqnarray}}
%A8
{\setlength\arraycolsep{1pt}
\begin{eqnarray}
c_{0}&=& \frac{1}{16} \left( Y_{2}+4Y_{1}+3Y_{0} \right) \nonumber \\
c_{1} & = & \frac{1}{2} \left(X_{2}+2 X_{1} \right) \nonumber \\
c_{2} &=& \frac{3}{2} \left( Y_{2} -Y_{0} \right) \\
c_{3} &=& 2\left( X_{2}-2X_{1} \right) \nonumber \\
c_{4} &=& Y_{2} -4Y_{1}+3Y_{0}  \nonumber
\end{eqnarray}}

\noindent $z=6$ case:
%A9
{\setlength\arraycolsep{1pt}
\begin{eqnarray}
a_{0}&=& \frac{1}{128} \left( F_{3}+6F_{2}+15F_{1}+10F_{0} \right) \nonumber \\
a_{1} & = & \frac{3}{32} \left(G_{3}+4G_{2}+5 G_{1} \right) \nonumber \\
a_{2} &=& \frac{15}{32} \left( F_{3}+2F_{2}-F_{1} -2F_{0} \right)\nonumber \\
a_{3} &=& \frac{5}{4}\left( G_{3}-3G_{1} \right)  \\
a_{4} &=& \frac{15}{8} \left(F_{3}- 2 F_{2} -F_{1}+2F_{0} \right) \nonumber \\
a_{5} &=& \frac{3}{2} \left( G_{3}-4G_{2}+5G_{1} \right) \nonumber\\
a_{6} &=& \frac{1}{2} \left( F_{3}-6 F_{2} +15F_{1}-10F_{0} \right) \nonumber
\end{eqnarray}}
%A10
{\setlength\arraycolsep{1pt}
\begin{eqnarray}
b_{0}&=& \frac{1}{256} \left( G_{3}+4G_{2}+5G_{1} \right) \nonumber \\
b_{1} & = & \frac{1}{64} \left(3F_{3}+8F_{2}+5F_{1} \right) \nonumber \\
b_{2} &=& \frac{5}{64} \left( 3G_{3}+4G_{2}-G_{1}  \right)\nonumber \\
b_{3} &=& \frac{5}{8}\left( F_{3}-F_{1} \right)  \\
b_{4} &=& \frac{5}{16} \left(3G_{3}- 4G_{2} -G_{1} \right) \nonumber \\
b_{5} &=& \frac{1}{4} \left( 3F_{3}-8F_{2}+5F_{1} \right) \nonumber\\
b_{6} &=& \frac{1}{4} \left( G_{3}- 4G_{2} +5G_{1} \right) \nonumber
\end{eqnarray}}
%A11
{\setlength\arraycolsep{1pt}
\begin{eqnarray}
c_{0}&=& \frac{1}{64} \left( Y_{3}+6Y_{2}+15Y_{1}+10Y_{0} \right) \nonumber \\
c_{1} & = & \frac{3}{16} \left(X_{3}+4X_{2}+5X_{1} \right) \nonumber \\
c_{2} &=& \frac{15}{16} \left( Y_{3}+2Y_{2}-Y_{1} -2Y_{0}  \right)\nonumber \\
c_{3} &=& \frac{5}{2}\left( X_{3}-3X_{1} \right)  \\
c_{4} &=& \frac{15}{4} \left(Y_{3}- 2Y_{2} -Y_{1} +2Y_{0} \right) \nonumber \\
c_{5} &=& 3 \left( X_{3}-4X_{2}+5X_{1} \right) \nonumber\\
c_{6} &=& Y_{3}- 6Y_{2} + 15Y_{1} -10Y_{0} \nonumber
\end{eqnarray}}

\end{document}